\documentclass{PoS}

\title{Supermassive and Intermediate-Mass Black Hole Growth at Galaxy Centers and Resulting Feedback using Cosmological Simulations} 

\ShortTitle{Massive Central BH Growth and Feedback in Galaxies} 

\author{\speaker{Paramita Barai} \\ 
	Instituto de Astronomia, Geof\'isica e Ci\^encias Atmosf\'ericas - 
	Universidade de S\~ao Paulo (IAG-USP), Rua do Mat\~ao 1226, S\~ao Paulo, 05508-090, Brasil \\
        E-mail: \email{paramita.barai@iag.usp.br}}

\author{Elisabete M. de Gouveia Dal Pino \\
        Instituto de Astronomia, Geof\'isica e Ci\^encias Atmosf\'ericas - 
        Universidade de S\~ao Paulo (IAG-USP), Rua do Mat\~ao 1226, S\~ao Paulo, 05508-090, Brasil \\
        E-mail: \email{dalpino@iag.usp.br}}

\abstract{
Accretion of matter onto central Black Holes (BHs) in galaxies liberates enormous amounts of feedback energy,
which influence the formation and evolution of structures, affecting the environment from pc to Mpc scales.
These BHs are usually Supermassive BHs (SMBHs: mass $\geq 10^6 M_{\odot}$) 
existing at the centers of active galactic nuclei (AGN),
which are widely observed through their multi-wavelength emission at all cosmic epochs.
The SMBH energy output is often observed as powerful AGN outflows in a wide variety of forms.
Relatively recently, Intermediate-Mass BHs (IMBHs: mass = $100 - 10^6 M_{\odot}$) 
have started to be observed hosted in Dwarf Galaxy (DG) centers.
Some of the central IMBHs in DGs show signatures of activity in the form of low-luminosity AGN. 

We have performed Cosmological Hydrodynamical Simulations to probe SMBHs in high-z quasars \cite{Barai17},
and IMBHs in DGs \cite{BaraiPino18}.
Our simulations employ the 3D TreePM SPH code GADGET-3, and include metal cooling, star formation,
chemical enrichment, stellar evolution, supernova feedback, AGN accretion and feedback.
Analyzing the simulation output in post-processing, we investigate the growth of the first IMBHs,
and the growth of the first SMBHs, their impact on star-formation, 
as well as their co-evolution with the respective host galaxies.
We quantify the impact of SMBHs and IMBHs on their host galaxies,
especially the effects on quenching star-formation.
We also study the corresponding BH outflow properties. 
} 

\FullConference{International Conference on Black Holes as Cosmic Batteries: UHECRs and Multimessenger Astronomy - BHCB2018\\
		12-15 September, 2018\\
		Foz du Iguazu, Brasil}

\begin{document}

\section{Introduction}

Active galactic nuclei (AGN) emit enormous amounts of energy powered by
the accretion of gas onto their central supermassive black holes (SMBHs) (e.g., \cite{rees84}).
Feedback from AGN are believed to strongly influence the formation
and evolution of galaxies (e.g., \cite{richstone98, Barai08}).
A strong manifestation of AGN feedback are AGN outflows observed
in a wide variety of forms (e.g., \cite{Crenshaw03, Feruglio15}). 

Quasars are very powerful AGN existing more commonly at high-$z$
than in the local Universe (e.g., \cite{Fan06}).
In the host galaxy of the quasar SDSS J1148+5251 at $z = 6.4$,
\cite{Maiolino12} detected broad wings of the [CII] line tracing
a massive outflow with velocities up to $\pm 1300$ km/s.
Follow-up by \cite{Cicone15} revised the mass outflow rate
lower limit to $1400 M_{\odot}$/yr.
The physical mechanisms by which quasar outflows affect their
host galaxies remain as open questions.

SMBHs of mass $\geq 10^9 M_{\odot}$ are observed to be in place
in luminous quasars by $z \sim 6$, when the Universe was less than $1$ Gyr
old (e.g., \cite{Wu15}).
It is difficult to understand how these early SMBHs formed over such short
time-scales, and there are open issues with various plausible scenarios
(e.g., \cite{Matsumoto15}).
AGN feedback should operate mostly in the negative form quenching star formation,
as suggested by observations (e.g., \cite{Schawinski06}),
and simulations (e.g., \cite{Scannapieco05}).
At the same time, AGN feedback can occasionally be positive,
by inducing star-formation, as have been shown in theoretical and numerical studies (e.g., \cite{Zubovas13}),
and observed in jet-induced star formation and radio-optical alignment (e.g., \cite{Zinn13}). 

Black holes are usually observed to belong to two populations: stellar-mass $(M_{\rm BH} \leq 10 - 100 M_{\odot})$ BHs, 
and supermassive $(M_{\rm BH} \geq 10^{6} M_{\odot})$ BHs. 
By natural extension, there should be a population of Intermediate-Mass Black Holes
(IMBHs: with mass between $100 - 10^6 M_{\odot}$) in the Universe. 
Analogous to SMBHs producing AGN feedback, the IMBHs should also have feedback. 

AGN feedback mechanism has recently started to been observed in low-mass galaxies.
Investigating the presence of AGN in nearby dwarf galaxies
using mid-infrared emission, \cite{Marleau17} identified $303$ candidates,
of which $91 \%$ were subsequently confirmed as AGN by other methods.
The stellar masses of these galaxies are estimated to be between $10^{6} - 10^{9} M_{\odot}$;
and the black hole masses in the range $10^{3} - 10^{6} M_{\odot}$.
\cite{Penny17} presented observational evidence for AGN feedback in a sample of
$69$ quenched low-mass galaxies $(M_{\star} < 4 \times 10^{9} M_{\odot})$;
including $6$ galaxies showing signatures of an active AGN preventing ongoing star-formation. 

The concordance $\Lambda$CDM cosmological scenario of galaxy formation presents 
multiple challenges in the dwarf galaxy mass range: e.g. core versus cusp density profile, number of DGs. 
Recently \cite{Silk17} made an exciting claim that the presence of IMBHs 
at the centers of essentially all old Dwarf Galaxies (DGs) can potentially solve the problems. 
Early feedback from these IMBHs output energy and affect the host gas-rich DGs at $z = 5 - 8$. 
This early feedback can quench star-formation, reduce the number of DGs, 
and impact the density profile at DG centers. 

In this work we present results of the growth and feedback of SMBHs in AGN, that of and IMBHs in DGs. 
We focus on negative BH feedback effects where star-formation is quenched. 
We performed zoomed-in cosmological hydrodynamical simulations 
of quasar-host galaxies at $z \geq 6$ to study their outflows (details in \cite{Barai17}). 
Our goals are to investigate the impact of AGN outflows on host galaxies
in the early Universe, and compute the simulated outflow properties;
for which both observations and theoretical studies are scarce. 
We also performed cosmological hydrodynamical simulations of periodic 
comoving $(2 Mpc)^3$ volumes, starting from $z=100$, to study IMBHs (details in \cite{BaraiPino18}). 
Here, we investigate the scenario that IMBHs are present at the centers of all dwarf galaxies.
Our goals are to test if IMBHs would grow at DG centers, and quantify the impact on star formation.

\section{Numerical Method and Simulations}
\label{sec-numerical}

The initial conditions at $z = 100$ are generated using the
{\sc MUSIC}\footnote{MUSIC - Multi-scale Initial Conditions for Cosmological Simulations: https://bitbucket.org/ohahn/music} software \cite{Hahn11}. 
We use a modified version of the TreePM (particle mesh) - SPH (smoothed particle hydrodynamics) 
code {\sc GADGET-3} (\cite{Springel05}) to perform our cosmological hydrodynamical simulations. 
Radiative cooling and heating is incorporated from \cite{Wiersma09a}.
Eleven element species (H, He, C, Ca, O, N, Ne, Mg, S, Si, Fe) are tracked.
Star-formation is implemented following the multiphase effective sub-resolution
model by \cite{SH03}, and chemical enrichment from \cite{Tornatore07}.

BHs are collisionless sink particles (of mass $M_{\rm BH}$) in our simulations.
A BH (of initial mass $M_{\rm BHseed}$) is seeded at the center of each galaxy more
massive than a total mass $M_{\rm HaloMin}$, which does not contain a BH already.
We test different values of minimum halo mass and seed BH mass in the range:
$M_{\rm HaloMin} = (10^{6} - 10^{7}) M_{\odot}$,
and $M_{\rm BHseed} = (10^{2} - 10^{3}) M_{\odot}$.
The sub-resolution prescriptions for gas accretion onto BHs and 
{\it kinetic} feedback are adopted from \cite{Barai14, Barai16}. 
The halo mass $(M_{\rm halo})$ of a galaxy, and its virial radius in comoving coordinates $(R_{200})$, are related 
such that $R_{200}$ encloses a density $200$ times the mean comoving matter density of the Universe: 
\begin{equation} 
\label{eq-Mhalo} 
M_{\rm halo} = \frac{4 \pi}{3} R_{200}^3 \left(200 \rho_{\rm crit} \Omega_{M,0}\right) , 
\end{equation} 
where $\rho_{\rm crit} = 3 H_0^2 / (8 \pi G)$ is the present critical density. 

We execute a series of $4$ Zoomed-In cosmological hydrodynamical simulations, 
with characteristics listed in Table~\ref{Table-Sims-1}.
All the $4$ runs incorporate metal cooling, chemical enrichment, SF and SN feedback.
The first run has no AGN included,
while the latter three explore different AGN feedback models. 
To create the zoom-in region, first a dark-matter (DM) only low-resolution simulation is carried out
of a $(500 ~ {\rm Mpc})^3$ comoving volume, using $256^3$ DM particles,
from $z = 100$ up to $z = 6$.
Halos are identified within it using the {\it Friends-of-Friends} (FOF) algorithm.
We select the most-massive halo at $z = 6$, of a total mass $M_{\rm halo} = 4.4 \times 10^{12} M_{\odot}$,
and a virial radius $R_{200} \simeq 511$ kpc comoving.
We select the DM particles around it, inside a cubic box of side $2 R_{200}$.
These DM particles are tracked back to our initial condition at $z=100$,
and the Lagrangian region occupied by them is determined.
The Lagrangian region (volume of $(5.21 ~ {\rm Mpc})^3$) is populated with
particles of higher resolution: DM and baryons, going from $8$ to $13$ levels of refinement. 
Finally, we restart the zoom-in cosmological simulation. 
The high-resolution particle masses are: $m_{\rm DM} = 7.54 \times 10^{6} M_{\odot}$,
and $m_{\rm gas} = 1.41 \times 10^{6} M_{\odot}$. 
We employ $L_{\rm soft} = 1 /h$ kpc comoving as the Plummer-equivalent softening length 
for gravitational forces, for these high-resolution particles. 



\begin{table*}
\begin{minipage}{1.0 \linewidth}
\caption{Zoomed-In Cosmological Hydrodynamical Simulations (for SMBHs)}
\label{Table-Sims-1}
\begin{tabular}{@{}ccccc}

\hline

Run & AGN feedback & Reposition of BH & Geometry of region where & Half opening angle \\
name & algorithm & to potential-minimum & feedback is distributed & of effective cone \\

\hline

{\it noAGN} & No BH & -- & -- & -- \\   

{\it AGNoffset} & Kinetic & No & Bi-Cone & $45^{\circ}$ \\   

{\it AGNcone} & Kinetic & Yes & Bi-Cone & $45^{\circ}$ \\   

{\it AGNsphere} & Kinetic & Yes & Sphere & $90^{\circ}$ \\   

\hline
\end{tabular}

\end{minipage}
\end{table*}



\begin{table}
\begin{minipage}{1.0 \linewidth}
\caption{Periodic-Box Cosmological Hydrodynamical Simulations (for IMBHs)} 
\label{Table-Sims-2}
\begin{tabular}{@{}cccccc}

\hline

Run  & BH      & Min. Halo Mass for BH Seeding, & Seed BH Mass,                & BH kinetic feedback \\
name & present & $M_{\rm HaloMin} [M_{\odot}]$  & $M_{\rm BHseed} [M_{\odot}]$ & kick velocity $v_w$ (km/s) \\

\hline

{\it SN}         & No  & -- & -- & -- & \\                               

{\it BHs2h1e6}   & Yes & $h^{-1} \times 10^{6}$ & $10^{2}$ & $2000$ \\   

{\it BHs2h7e7}   & Yes & $5 h^{-1} \times 10^{7}$ & $10^{2}$ & $2000$ \\ 

{\it BHs3h1e7}   & Yes & $1 \times 10^{7}$ & $10^{3}$ & $2000$ \\        

{\it BHs3h2e7}   & Yes & $2 \times 10^{7}$ & $10^{3}$ & $2000$ \\        

{\it BHs3h3e7}   & Yes & $3 \times 10^{7}$ & $10^{3}$ & $2000$ \\        

{\it BHs3h4e7}   & Yes & $4 \times 10^{7}$ & $10^{3}$ & $2000$ \\        

{\it BHs3h4e7v5} & Yes & $4 \times 10^{7}$ & $10^{3}$ & $5000$ \\        

{\it BHs3h5e7}   & Yes & $5 \times 10^{7}$ & $10^{3}$ & $2000$ \\        

{\it BHs4h4e7}   & Yes & $4 \times 10^{7}$ & $10^{4}$ & $2000$ \\        

\hline
\end{tabular}

\end{minipage}
\end{table}


We perform cosmological hydrodynamical simulations of small-sized 
boxes with periodic boundary conditions, to probe dwarf galaxies at high redshifts. 
We execute a series of $10$ simulations, with characteristics listed in Table~\ref{Table-Sims-2}. 
The size of the cubic cosmological volume is $(2 h^{-1} ~ {\rm Mpc})^3$ comoving, and 
we use $256^3$ dark matter and $256^3$ gas particles. 
The dark matter particle mass is $m_{\rm DM} = 3.44 \times 10^{4} h^{-1} M_{\odot}$,
and the gas particle mass is $m_{\rm gas} = 6.43 \times 10^{3} h^{-1} M_{\odot}$.
The gravitational softening length is set to $L_{\rm soft} = 0.1 h^{-1}$ kpc comoving.
Black holes of mass $1000 M_{\odot}$ are seeded inside
halos when they reach a mass of $10^7 M_{\odot}$.
The black holes grow by accretion of gas from their surroundings
and by merger with other black holes, and consequently eject feedback energy.

\section{Results and Discussion}
\label{sec-results}

\subsection{Black Hole Accretion and Growth}
\label{sec-res-BH-Growth}

\begin{figure}
\centering
\includegraphics[width = 1.05 \linewidth]{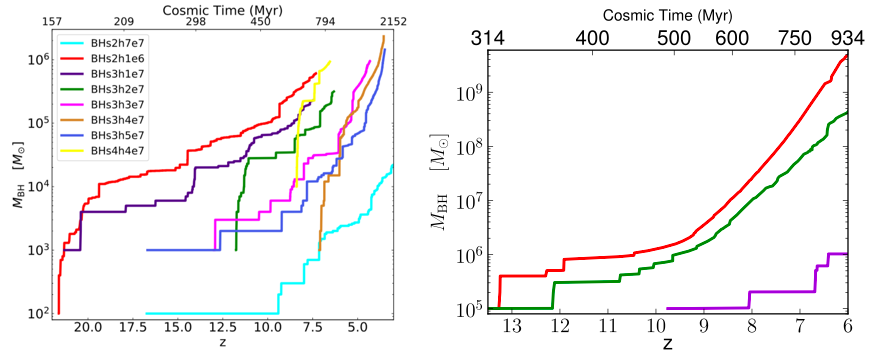}
\caption{ 
BH mass growth with redshift of the most-massive BH in each run. The different colours discriminate the various runs. 
Left panel: Periodic-Box cosmological simulations showing growth of IMBHs. (Figure modified from \cite{BaraiPino18}).
Right panel: Zoomed-In cosmological simulations showing growth of SMBHs. (Figure modified from \cite{Barai17}). 
} 
\label{fig-BH-Mass-vs-Time}
\end{figure} 

We find that first BHs are seeded at different cosmic times depending on the value of
minimum halo mass for BH seeding, $M_{\rm HaloMin}$.
The seeding epoch varies between $z \sim 22$ to $z \sim 16$ in our periodic-box cosmological simulations, when the 
first halos reach $M_{\rm halo} = h^{-1} \times 10^{6} M_{\odot}$ to $M_{\rm halo} = 5 \times 10^{7} M_{\odot}$.
The redshift evolution of the most-massive BH mass in these periodic-box simulation runs 
is plotted in {\it Fig.~\ref{fig-BH-Mass-vs-Time} - left panel}.
Each BH starts from an initial seed of $M_{\rm BH} = 10^{2} M_{\odot}$ in the runs named {\it BHs2*},
$10^{3} M_{\odot}$ in the runs named {\it BHs3*}, and $10^{4} M_{\odot}$ in the runs named {\it BHs4*}.
The subsequent mass growth is due to merger with other BHs (revealed as vertical rises in $M_{\rm BH}$),
and gas accretion (visualized as the positive-sloped regions of the $M_{\rm BH}$ versus $z$ curve).
The final properties reached depends on the simulation.
The most-massive BH, considering all the runs, has grown to $M_{\rm BH} = 2 \times 10^6 M_{\odot}$
at $z = 5$ in run {\it BHs3h4e7} (brown curve in Fig.~\ref{fig-BH-Mass-vs-Time} - left panel). 

The redshift evolution of the most-massive BH mass in the three AGN runs of the zoomed-in cosmological 
simulations is plotted in {\it Fig.~\ref{fig-BH-Mass-vs-Time} - right panel}. 
Each BH starts as a seed of $M_{\rm BH} = 10^5 M_{\odot}$, 
at $z \sim 14$ in the runs {\it AGNcone} and {\it AGNsphere} ($z \sim 10$ in {\it AGNoffset}).
The subsequent growth is due to merger with other BHs and gas accretion. 
The dominant mode of BH growth occurs over the redshifts $z = 9 - 6$
in runs {\it AGNcone} and {\it AGNsphere},
corresponding to Eddington-limited gas accretion where Eddington ratio $= 1$.
The $\dot{M}_{\rm BH}$ has a power-law increase,
and the BH mass increases by a factor $\sim 10^3$.
The final properties reached at $z = 6$ depends on the simulation;
e.g. $M_{\rm BH} = 4 \times 10^9 M_{\odot}$
and $\dot{M}_{\rm BH} = 100 M_{\odot}$/yr in run {\it AGNcone} (red curve).
There is variability of the $\dot{M}_{\rm BH}$,
whereby it fluctuates by a factor of up to $100$.
The BH grows 10 times more massive at $z = 6$ in the {\it AGNcone} case
than in the {\it AGNsphere} run.
This is because more gas can inflow along the perpendicular direction
to the bi-cone, and accrete onto the BH.

\subsection{Star Formation}
\label{sec-res-SF}

\begin{figure}
\centering
\includegraphics[width = 1.05 \linewidth]{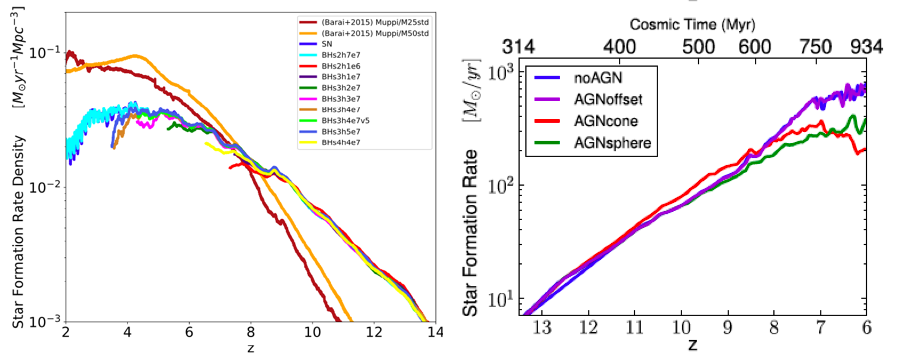}
\caption{ 
Left panel: Total star formation rate density (in $M_{\odot} yr^{-1} Mpc^{-3}$, total SFR integrated over simulation 
volume) as a function of redshift, in the Periodic-Box cosmological simulations. (Figure modified from \cite{BaraiPino18}). 
Right panel: Sum total star formation rate (in $M_{\odot} yr^{-1}$) as a function of 
redshift, in the Zoomed-In cosmological simulations. (Figure modified from \cite{Barai17}). 
}
\label{fig-SFR-total}
\end{figure} 

Stars form in the simulation volume from cold dense gas. 
The Star Formation Rate Density
(SFRD in units of $M_{\odot} yr^{-1} Mpc^{-3}$, counting stars forming in the whole simulation box)
versus redshift of the periodic-box cosmological simulation runs is displayed in {\it Fig.~\ref{fig-SFR-total} - left panel}. 
The SFRD rises with time in the {\it SN} run (blue curve in Fig.~\ref{fig-SFR-total} - left panel) initially from $z \sim 15$,
reaches a peak at $z \sim 4$ (the peak epoch of star-formation activity in the Universe),
and decreases subsequently over $z \sim 4 - 2$.
The presence of a BH quenches star formation by accreting some gas in,
ejecting some gas out of the halo as outflows, and/or heating the gas.
The models suppress SF substantially from $z \sim 8$ onwards, when the BHs have grown massive.
We find that BHs need to grow to $M_{\rm BH} > 10^5 M_{\odot}$,
in order to suppress star-formation, even in these dwarf galaxies.
BH feedback causes a reduction of SFR up to $5$ times in the different runs.

The red curve (run {\it BHs2h1e6}) already quenches SF as early as $z \sim 8$.
This is because the BH has already grown to $M_{\rm BH} \sim 5 \times 10^{5} M_{\odot}$ at that epoch,
more massive than all the other runs.
As another example, the brown (run {\it BHs3h4e7}) and royal-blue (run {\it BHs3h5e7}) curves
quench SF from $z \sim 4.5$ to $z \sim 3.5$.
This is the epoch when the BH masses in these runs increase from
$M_{\rm BH} = 10^{5} M_{\odot}$ to $M_{\rm BH} = 10^{6} M_{\odot}$ 
(as can be seen from Fig.~\ref{fig-BH-Mass-vs-Time} - left panel). 

The star formation rate (total SFR in the simulation box) versus redshift of the four zoomed-in cosmological 
simulations is displayed in {\it Fig.~\ref{fig-SFR-total} - right panel}. 
The SFR rises with time in all the runs initially,
and continues to increase in the {\it noAGN} case without a BH.
The SFR in run {\it AGNoffset} is almost similar to that in the run {\it noAGN},
because the BHs are too small there to generate enough feedback.
A similar outcome happens in the runs {\it AGNcone} and {\it AGNsphere}
at $z \geq 8$, when the BHs are too small.

The models suppress SF substantially from $z \sim 8$ onwards,
when the BHs have grown massive and generate larger feedback energy.
Thus, we find that BHs need to grow to $M_{\rm BH} > 10^7 M_{\odot}$,
in order to suppress star-formation, even in massive galaxies
(of $M_{\star} = 4 \times 10^{10} M_{\odot}$, and specific-SFR $= 5 \times 10^{-9}$ yr$^{-1}$).
BH feedback causes a reduction of SFR up to $4$ times at $z = 6$:
from $800 M_{\odot}$/yr in the {\it noAGN} run, to $200 M_{\odot}$/yr in run {\it AGNcone}, 
and $350 M_{\odot}$/yr in run {\it AGNsphere}.

\subsection{Large Scale Environment of IMBHs}
\label{sec-res-LargeScale}

\begin{figure*}
\centering
\includegraphics[width = 1.07 \linewidth]{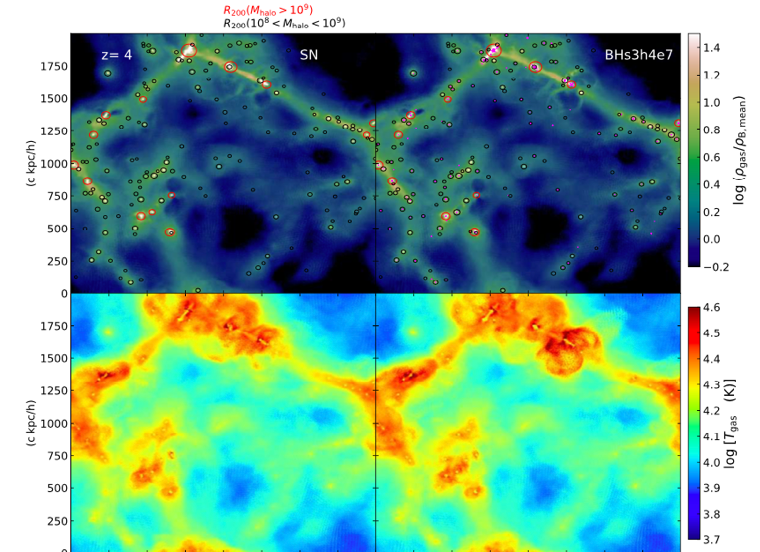} 
\caption{ 
Projected gas kinematics in the whole $(2000 h^{-1}$ kpc$)^3$ periodic cosmological simulation box at $z = 4$. 
The two rows present the gas overdensity (top row), and gas temperature (bottom row). 
The two columns are for different periodic-box simulations: {\it SN} (left) and {\it BHs3h4e7v2} (right). 
The red circles in the top row depict the virial radius $R_{\rm 200}$ of galaxies 
in the mass range $M_{\rm halo} > 10^{9} M_{\odot}$, while the black circles 
show the $R_{\rm 200}$ of $10^{8} < M_{\rm halo} < 10^{9} M_{\odot}$ galaxies. 
The positions of BHs in run {\it BHs3h4e7v2} are indicated by the magenta cross-points in the top-right panel. 
(Figure modified from \cite{BaraiPino18}). 
} 
\label{fig-Gas-rho-T-SFR}
\end{figure*}

The gas morphology in our periodic cosmological simulation box, or the large scale structures,
is plotted in {\it Fig.~\ref{fig-Gas-rho-T-SFR}}.
It displays the projected gas kinematics in the whole $(2000 h^{-1}$ kpc$)^3$
comoving volume at $z = 4$, for two periodic-box simulations:
{\it SN} (left column) and {\it BHs3h4e7v2} (right column).
The overdensity (i.e., the ratio between the gas density and the cosmological mean baryon density in the Universe) 
and temperature of the gas are plotted in the two rows from the top.
The black circles in the top row depict the virial radius $R_{\rm 200}$
(defined in Eq.~(\ref{eq-Mhalo})) of dwarf galaxies with halo masses in the range
$10^{8} \leq M_{\rm halo} \leq 10^{9} M_{\odot}$.
The red circles show the $R_{\rm 200}$ of relatively massive galaxies, those
having higher halo masses $M_{\rm halo} > 10^{9} M_{\odot}$.

The spatial locations of the BHs within our {\it BHs3h4e7v2} simulation box
can be visualized in the top-right panel of Fig.~\ref{fig-Gas-rho-T-SFR}.
Here the magenta cross-points designate BH positions,
overplotted with the gas overdensity. In this run, BHs are seeded at the
centres of galaxies with $M_{\rm HaloMin} = 4 \times 10^{7} M_{\odot}$.
Therefore all the red circles ($M_{\rm halo} > 10^{9} M_{\odot}$ galaxies)
and most of the black circles ($M_{\rm halo} = 10^{8}-10^{9} M_{\odot}$
galaxies) contain BHs at their centres. 

The cosmological large-scale-structure filaments are visible in all the panels of both the runs.
There are three Mpc-scale filaments: extending from east to north, from west to north, and from west to south.
In addition, there is an overdense region running from the center of the box to the south-west.
The filaments consist of dense (yellow and white regions in the top panels), and star-forming gas.
The massive galaxies (red circles) lie at the high-density intersections of the filaments, or in the filaments.

In terms of temperature, the immediate vicinity of the dense filaments consists of
hotter gas ($T \sim 10^{4.6}$ K, red regions in the bottom panels of
Fig.~\ref{fig-Gas-rho-T-SFR}), as compared to that in the low-density
intergalactic medium and voids (yellow and blue regions).
Several mechanisms play together to heat the gas to higher temperatures
in the filament vicinity.
There are global environmental processes like shock heating during
galaxy mergers, and large-scale-structure formation,
which are present in both the {\it SN} and {\it BHs3h4e7v2} runs.
Acting together there are local galactic processes like feedback driven by SN (present in both the columns),
and BHs (present in the right column only), which heats the gas, and also often generate outflows.

As they accrete and grow, the BHs provide feedback energy (according to the prescription described in \cite{BaraiPino18}),
which may drive gas outflows.
High-velocity gas ejected by central BH feedback propagates radially outward,
and shocks with the surrounding slower-moving gas, creating bubble-like gas outflows.
Our simulation {\it BHs3h4e7v2} shows the formation of BH feedback-induced outflows,
as can be seen in Fig.~\ref{fig-Gas-rho-T-SFR} in the top-right half
of the right column panels, around the most-massive BH.
The outflows are extended bipolar oval-shaped regions along
the north-east to south-west direction, propagating to about $10 \times R_{\rm 200}$.
The outflows consist of hot ($T > 10^{4.6}$ K) - visible as red areas
in the temperature map (bottom-right panel), and low-density gas (top-right panel).

\subsection{Quasar Outflow Morphology around SMBHs}
\label{sec-res-outflow-Morph}

\begin{figure*}
\centering
\includegraphics[width = 1.0 \linewidth]{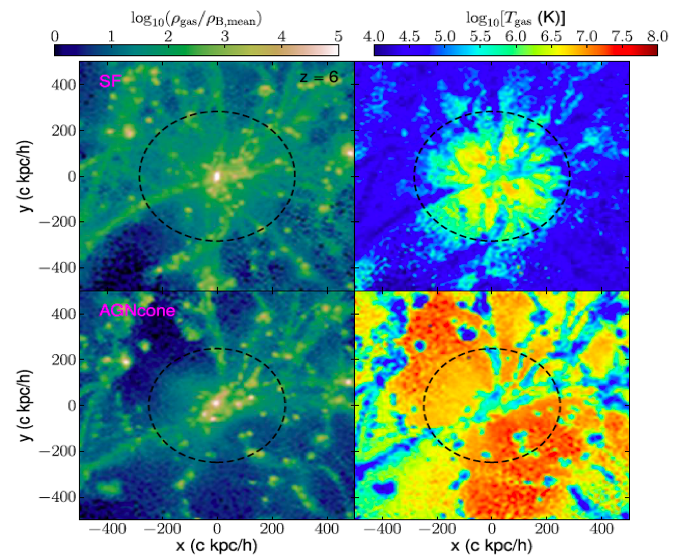}
\caption{
Projected gas kinematics in a $(1000 h^{-1}$ kpc$)^3$ comoving volume around the most-massive galaxy at $z = 6$, 
in a Zoomed-In cosmological simulation. 
The two columns indicate gas overdensity (left column), and gas temperature (right column). 
The two rows are for different simulations: {\it SF} (top row), and {\it AGNcone} (bottom row). 
The black dashed circle depicts the galaxy virial radius $R_{\rm 200}$ in each case. (Figure modified from \cite{Barai17}). 
} 
\label{fig-Gas-rho-T-SFR-velR}
\end{figure*}

We find prominent gas outflows being developed in our simulated massive quasar-host galaxies, 
in the zoomed-in cosmological volumes. 
While the SMBHs output feedback energy, 
high-velocity gas propagates radially outward and shocks with the surrounding slower-moving gas. 
This creates bubble-like gas outflows originating from the central SMBH. 
The outflow morphology is plotted in {\it Fig.~\ref{fig-Gas-rho-T-SFR-velR}}, which displays the projected gas 
kinematics in a $(1000 h^{-1}$ kpc$)^3$ comoving volume in two zoom-in cosmological simulation runs, at $z = 6$. 
The overdensity and temperature of the gas is plotted in the two columns. 

In the {\it noAGN} run (top row), the outflows are weak ($v_{r} < 300$ km/s), and 
bounded within $0.5 R_{\rm 200}$ as warm-hot ($T \sim 10^6$ K) halo gas. 
It is caused by SN feedback and galaxy merger shocks. 
Such a high-mass $(M_{\star} \sim 10^{11} M_{\odot})$ galaxy cannot efficiently drive strong outflows 
with only SN feedback (e.g., \cite{Benson03, Talia16}). 

The other two simulations ({\it AGNcone} and {\it AGNsphere}) show the formation of BH feedback-induced 
powerful outflows, which are hot ($T \sim 10^8$ K) - visible as yellow and red areas in the temperature plot 
(bottom-right panel of Fig.~\ref{fig-Gas-rho-T-SFR-velR}), and consist of low-density, metal-enriched gas.
The outflows are fastest ($v_{r} > 2000$ km/s) in run {\it AGNcone}
(bottom row, where the BHs become more-massive than the other runs), extended bipolar shaped,
propagating to beyond the galaxy $R_{\rm 200}$ (black dashed circle). 
The outflows disrupt the cold dense filamentary gas inflows to the galaxy center, along the direction of outflow propagation.
This quenches star formation, and halts the formation of nearby satellite galaxies.
It is revealed by the disrupted clumps and lack of star-forming dense filaments
near the central galaxy in runs {\it AGNcone} and {\it AGNsphere}, as compared to run {\it noAGN}.
The outflows transport metals away from star-forming regions,
and enrich the surrounding circumgalactic medium out to $R_{\rm 200}$.
Some inflows of cold, dense gas continue to occur perpendicular to the outflow direction. 

We find that the density increment, at the edges of the outflow shocks, 
remains below the SF threshold density parameter (\S\ref{sec-numerical}) of our simulations. 
Therefore no new star-formation is triggered in our simulations by AGN feedback. 
This might be a numerical resolution effect, since our resolution (length scale of $1$ kpc comoving) 
is lower than those where positive feedback from AGN is simulated in galaxies (e.g., \cite{Bieri15}).

\section{Conclusions} 

Intermediate-mass black holes (mass between $100 - 10^6 M_{\odot}$) 
have started to been observed at the centers of dwarf galaxies. 
We perform cosmological hydrodynamical simulations of $(2 h^{-1} ~ {\rm Mpc})^3$ comoving boxes 
with periodic boundary conditions, to probe dwarf galaxies and central IMBHs at high redshifts. 
We conclude that IMBHs at DG centers grow from $10^{2} - 10^{3} M_{\odot}$ 
to $10^{5} - 10^{6} M_{\odot}$ by $z \sim 4$ in a cosmological environment.
These IMBHs in DGs can become the seeds of supermassive BHs
(which grows to $M_{\rm BH} \sim 10^9 M_{\odot}$) in massive galaxies. 
Star formation is quenched when the BHs have grown to $M_{\rm BH} > 10^5 M_{\odot}$.
We find a positive correlation between the mass growth BHs and the quenching of SF. 
Our conclusions, based on numerical simulation results, support the phenomenological ideas made by \cite{Silk17}. 
IMBHs at the centers of dwarf galaxies can be a strong source of feedback to 
quench star-formation and generate outflows. 
At the same time, these IMBHs form the missing link between stellar-mass and supermassive BHs. 

Gas accretion onto central supermassive black holes of active galaxies 
and resulting energy feedback, often manifested as AGN outflows, 
is an important component of galaxy evolution, 
whose details are still unknown especially at early cosmic epochs. 
We investigate outflows in quasar-host galaxies at $z \geq 6$ by performing cosmological hydrodynamical simulations. 
We simulate the $2 R_{200}$ region around a $2 \times 10^{12} M_{\odot}$ halo 
at $z = 6$, inside a $(500 ~ {\rm Mpc})^3$ comoving volume, using the zoomed-in technique. 
We find that, starting from $10^{5} M_{\odot}$ seeds BHs can grow to $10^{9} M_{\odot}$ in cosmological environments. 
During their growth, BHs accrete gas at the Eddington accretion rate over $z = 9 - 6$, for $100$s of Myr. 
At $z = 6$, our most-massive BH has grown to $M_{\rm BH} = 4 \times 10^9 M_{\odot}$. 
Fast ($v_{r} > 1000$ km/s), powerful ($\dot{M}_{\rm out} \sim 2000 M_{\odot}$/yr) 
outflows of shock-heated low-density gas form at $z \sim 7$, and propagate up to hundreds kpc. 
Star-formation is quenched over $z = 8 - 6$. 
The outflow mass is increased (and the inflow is reduced) by $\sim 20\%$.

\section{Acknowledgements} 

This work is supported by the Brazilian Funding Agencies 
FAPESP (grants 2016/01355-5, 2016/22183-8, and 2013/10559-5); 
and CNPq (grant 308643/2017-8).

\end{document}